\documentclass[twocolumn,prb,aps,psfig,showpacs,preprintnumbers,superscriptaddress]{revtex4}
\usepackage{graphicx}
\usepackage{epstopdf}
\usepackage{float}
\usepackage{color}
\usepackage{amsmath}

\begin{document}

\title{Electronic properties of GeTe and Ag- or Sb-substituted GeTe studied by low temperature $^{125}$Te NMR }

\author{J.~Cui}
\affiliation{Ames Laboratory, U.S. DOE, Ames, Iowa 50011, USA}
\affiliation{Department of Chemistry, Iowa State University, Ames, Iowa 50011, USA}
\author{E.~M.~Levin}
\affiliation{Ames Laboratory, U.S. DOE, Ames, Iowa 50011, USA}
\affiliation{Department of Physics and Astronomy, Iowa State University, Ames, Iowa 50011, USA}
\author{Y.~Lee}
\affiliation{Ames Laboratory, U.S. DOE, Ames, Iowa 50011, USA}
\author{Y.~Furukawa}
\affiliation{Ames Laboratory, U.S. DOE, Ames, Iowa 50011, USA}
\affiliation{Department of Physics and Astronomy, Iowa State University, Ames, Iowa 50011, USA}

\date{\today}

\begin{abstract}
    We have carried out $^{125}$Te nuclear magnetic resonance (NMR)  in a wide temperature range of 1.5 -- 300 K
to investigate electronic properties of Ge$_{50}$Te$_{50}$, Ag$_{2}$Ge$_{48}$Te$_{50}$ and Sb$_{2}$Ge$_{48}$Te$_{50}$
from a microscopic point of view.
    From the temperature dependence of NMR shift ($K$) and nuclear spin lattice relaxation rate (1/$T_1$), we found that two bands contribute to the physical properties of the materials.
    One band overlaps the Fermi level providing the metallic state where no strong electron correlations are revealed by Korringa analysis. 
      The other band is separated from the Fermi level by an energy gap of  $E_{\rm g}/k_{\rm B}$ $\sim$ 67 K, which gives rise to the semiconductor-like  properties.
     First principle calculation revealed that the metallic band originates from the Ge vacancy while the semiconductor-like band may be related to the fine structure of the density of states near the Fermi level. 
      Low temperature $^{125}$Te NMR data for the materials studied here  clearly show that the Ag substitution increases  hole concentration while Sb substitution decreases it.

\end{abstract}

\maketitle

 \section{Introduction}
    Complex tellurides have been studied extensively due to their intriguing fundamental properties and their application as thermoelectric materials,\cite{Hein1964, Steigmeier1970, Akola2008,Snyder2008,Heremans2008, Skarabek1995} which directly convert heat into electricity. 
    The efficiency is characterized by the dimensionless figure of merit $zT = S^2$$\sigma T/\kappa$ ($S$ the Seebeck coefficient, $\sigma$ the electrical conductivity, $T$ the temperature, and $\kappa$ the thermal conductivity).
    The well-known group of thermoelectric materials is complex tellurides based on GeTe,\cite{Levin2013, Wu2014, Sankar2015} TAGS-$m$ materials (GeTe)$_{m}$(AgSbTe$_2)_{100-m}$, having a thermoelectric figure of merit $zT$ above 1.\cite{Skarabek1995, Plachkova1984,Cook2007,Levin2012} 
   According to band calculations, GeTe  is a narrow band gap semiconductor whose band gap is calculated to be  0.3 $\sim$  0.5 eV.\cite{Herman1968, Cohen1968, Polatoglou1982, Singh2013}
    On the other hand, the electrical resistivity measurements show metallic behavior\cite{Levin2013, Damon1967, Kolomoets1, Kolomoets2, Gevacancy2006} although the small gap also has been observed by optical measurements.\cite{Nikolic1969}
    This is believed to be due to high hole concentrations generated by Ge vacancies, forming a self-dopant system with $p$-type conductivity. \cite{Levin2013,Damon1967,Lubell1963}
    Therefore, depending on the samples composition, they may have different concentrations of Ge vacancies resulting in different physical properties. 
   This makes it very difficult to understand physical properties of GeTe-based materials.
    In fact, there is a significant discrepancy between the electronic and thermal transport data for GeTe-based materials reported in the literature.\cite{Skarabek1995, Gelbstein2009, Gelbstein2010, Zhang2011} 

    In order to avoid such confusion, one would need to study the physical property using well characterized samples. 
   We have conducted systematic characterization of GeTe by using x ray diffraction (XRD), scanning electron microscopy (SEM), energy dispersive spectroscopy (EDS), Seebeck coefficient, electrical resistivity, Hall effect, thermal conductivity, and $^{125}$Te nuclear magnetic resonance (NMR) measurements.\cite{Levin2013}
   Hereafter we will use notation Ge$_{50}$Te$_{50}$ for GeTe with the coefficients shown in atomic percent.

      In our previous paper,\cite{Levin2013} we concluded that the discrepancy in the data for Ge$_{50}$Te$_{50}$ reported in literature can be attributed to the variation in the Ge/Te ratio of solidified samples as well as to different condition of measurements. 
    It is well established that NMR is a powerful tool to investigate carrier concentrations in semiconductors from a microscopic point of view. 
    It is noted that the Hall and Seebeck effects show only the bulk properties, which can be affected by small amounts of a second phase.\cite{Wolfe1960, Heremans2008_2}
    Nuclear spin lattice relaxation rates (1/$T_1$) have been measured at room temperature, and were found to increase linearly with carrier concentrations.\cite{Levin2013PRB} 
    However, to our knowledge, no systematic NMR investigation of Ge$_{50}$Te$_{50}$ has been carried out in a wide temperature range. 

   In this paper, we report the first $^{125}$Te NMR measurements of Ge$_{50}$Te$_{50}$  in a wide temperature range of $T$ = 1.5 - 300 K. 
    We found that the NMR shift $K$ and 1/$T_1T$ data are nearly temperature independent at low temperatures below $\sim$ 50 K and both increase slightly with increasing temperature at high temperatures. 
    These behavior can be well explained by  a two band model where one band overlaps the Fermi level and the other band is separated from the Fermi level by an energy gap of  $E_{\rm g}/k_{\rm B}$ $\sim$ 67 K.
    First principle calculations indicate that the first band originates from the Ge vacancy while the second band may be related to the fine structure of the density of states near the Fermi level. 
     We also carried out $^{125}$Te NMR measurements of M$_{2}$Ge$_{48}$Te$_{50}$ (M = Ag, Sb) to study carrier doping effects on electronic properties.
   Clear changes in carrier concentration by Ag or Sb substitutions were observed: the Ag substitution increases the hole concentration whereas Sb substitution decreases the concentration, which is consistent with our previous report.\cite{Levin2016}

 \section{Experimental}

    Polycrystalline samples of Ge$_{50}$Te$_{50}$, Ag$_{2}$Ge$_{48}$Te$_{50}$ and Sb$_{2}$Ge$_{48}$Te$_{50}$ were prepared by direct reaction of the constituent elements of Ge, Te, Ag or Sb in  fused silica ampoules, as described in Ref. \onlinecite{Levin2013} and Ref. \onlinecite{Levin2016}. 
    The samples were well characterized by XRD, Seebeck coefficient, electrical resistivity, Hall effects, and room temperature $^{125}$Te NMR  measurements. 
     The coarsely powdered samples were loosely packed into 6-mm quartz tube for NMR measurements.
      NMR measurements of $^{125}$Te ($I$ = $\frac{1}{2}$; $\frac{\gamma_{\rm N}}{2\pi}$ = 13.464 MHz/T) nucleus were conducted using a homemade phase-coherent spin-echo pulse spectrometer. 
   $^{125}$Te NMR spectra were obtained either by Fourier transform of the NMR echo signal at a constant magnetic field of 7.4089  T or by sweeping the magnetic field at a frequency of 99.6 MHz in the temperature range of $T$ = 1.5 - 300 K. 
    The NMR echo signal was obtained by means of a Hahn echo sequence with a typical $\pi$/2 pulse length of 7.5 $\mu$s.

 \section{Results and discussion}
  
\begin{figure}[t]
\centering
\includegraphics[width=8.5cm]{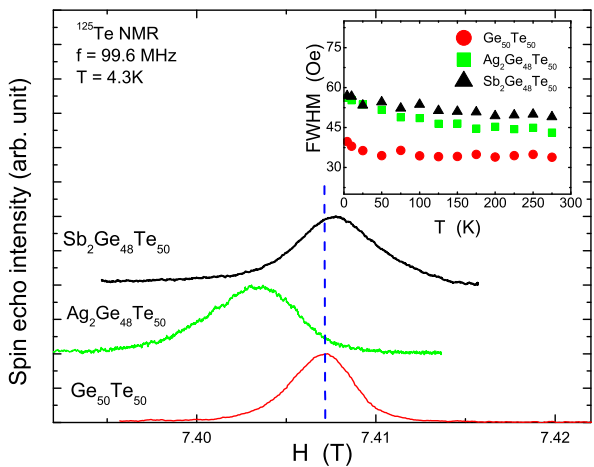}
\caption{(Color online) Field-swept $^{125}$Te-NMR spectra for Ge$_{50}$Te$_{50}$, Ag$_{2}$Ge$_{48}$Te$_{50}$, and 
Sb$_{2}$Ge$_{48}$Te$_{50}$ at $f$ = 99.6 MHz and $T$ = 4.3 K. 
The dotted vertical line is a guide for eyes.
The inset shows the temperature dependence of FWHM for the samples. }
\label{fig:spectrum}
\end{figure}


    Figure~\ref{fig:spectrum} shows field-swept $^{125}$Te NMR spectra measured at 4.3 K for Ge$_{50}$Te$_{50}$, Ag$_{2}$Ge$_{48}$Te$_{50}$ and Sb$_{2}$Ge$_{48}$Te$_{50}$. 
    The full width at half maximum (FWHM) for Ge$_{50}$Te$_{50}$ is 40.0(5)  Oe at $T$ = 4.3 K which  is almost independent of temperature although a slight increase can be observed below $\sim$ 25 K as shown in the inset of Fig. \ref{fig:spectrum}. 
   This FWHM is slightly smaller than 43 Oe at room temperature reported previously.\cite{Levin2016} 
    With Ag  substitution the peak position shifts to lower magnetic field, while the peak position slightly shifts to higher magnetic field with Sb substitution.
    The FWHM shows  a slight increase to 56.0(5) Oe and 54.0(5)  Oe at $T$ = 4.3 K  for Ag- or Sb-substituted samples, respectively. 
   The FWHM is also found to increase slightly with decreasing temperature for Ag- or Sb-substituted samples.
    These observed values are also closed to the values ($\sim$ 50 Oe) at room temperature reported previously.\cite{Levin2016} 
 
\begin{figure}[t]
\centering
\includegraphics[width=9.5cm]{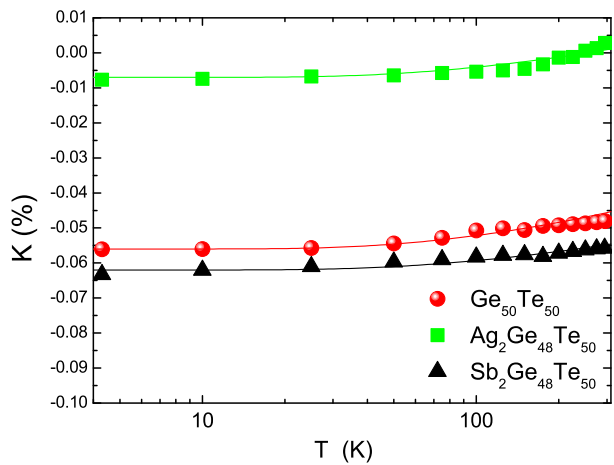}
\caption{(Color online) Temperature dependence of NMR shift $K$ for for Ge$_{50}$Te$_{50}$ (red circles), Ag$_{2}$Ge$_{48}$Te$_{50}$ (green squares) and Sb$_{2}$Ge$_{48}$Te$_{50}$ (black triangles). 
The solid lines are best fit using Eq.~\ref{eqn:KnightShift2}.}
\label{fig:Fig2}
\end{figure}

    The temperature dependence of NMR shift $K$ is shown in Fig.~\ref{fig:Fig2} where $K$ is determined by the peak position of the spectrum.
    Although the absolute values of $K$ depend on the samples composition, their temperature dependencies exhibit qualitatively the same behavior: $K$ slightly decreases with decreasing temperature, then levels off at low temperatures. 
   The temperature dependence of $K$ can be analyzed by a two band model where the first band overlaps the Fermi level and the second band is separated from the Fermi level by an energy gap ($E_{\rm g}$).
    The nearly temperature independent behavior observed at low temperatures is a typical characteristic for metal (due to Pauli paramagnetic susceptibility) originated from the first band.
    The increase of $K$ at  high temperatures originates from the second band, similar to the case of semiconductors. 
     Thus the total NMR shift is given by 
\begin{eqnarray}
K =   K_{\rm Pauli} +  K_{\rm semi} + K_{\rm orb}
\label{eqn:KnightShift}
\end{eqnarray}
where $K_{\rm Pauli}$ is the temperature independent NMR shift related to the the Pauli paramagnetic susceptibility $\chi_{\rm Pauli}$ due to self-doping/substitution effects and $K_{\rm semi}$ originates from the semiconducting-like  nature giving rise to the temperature dependent contribution because of thermal excitations across an energy gap $E_{\rm g}$.  
      The temperature independent $K_{\rm orb}$ includes chemical shift, orbital and Landau diamagnetic contributions. 
      As will be shown below, $K_{\rm orb}$ is estimated to be $-0.142$ $\%$.  
     As the temperature dependent $K_{\rm semi}$ has been calculated as $K_{\rm semi}$ $\propto$ $\sqrt{T}e^{-E_{\rm g}/k_{\rm B}T}$,\cite{Blembergen1954, Wolf1979}
the total $K$ is given as     
\begin{eqnarray}
K =   K_{\rm Pauli} + a \sqrt{T}e^{-E_{\rm g}/k_{\rm B}T} + K_{\rm orb}
\label{eqn:KnightShift2}
\end{eqnarray}
      Using the $K_{\rm orb}$ = $-0.142$ $\%$ and  $E_{\rm g}/k_{\rm B}$ = $67(4)$ K [$5.8(3)$ meV] estimated from the temperature dependence of 1/$T_1$ shown below, the experimental data are reasonably reproduced as shown by the solid lines with $ K_{\rm Pauli}$ = 0.084 $\%$,  $a$ = 0.00075 for Ge$_{50}$Te$_{50}$, $ K_{\rm Pauli}$ = 0.135 $\%$,  $a$ = 0.00057 for Ag$_2$Ge$_{48}$Te$_{50}$, and $ K_{\rm Pauli}$ = 0.081 $\%$,  $a$ = 0.00052 for Sb$_2$Ge$_{48}$Te$_{50}$, respectively.
     Since the $ K_{\rm Pauli}$ is proportional to Pauli paramagnetic susceptibility being proportional to the density of states ${\cal N}(E_{\rm F}$) at the Fermi level, the increase of $ K_{\rm Pauli}$  from Ge$_{50}$Te$_{50}$ to the Ag doped one indicates an increase of ${\cal N}(E_{\rm F}$) while Sb doping reduces ${\cal N}(E_{\rm F}$) at the Fermi level. 
    These results are consistent with the previous report.\cite{Levin2016}
     Note the ${\cal N}(E_{\rm F}$) discussed here is due to unavoidable self-doping and/or Ag(Sb)-substitution effects not including the effects of thermally activated carriers from the second band.

\begin{figure}[tb]
\centering
\includegraphics[width=8.8 cm]{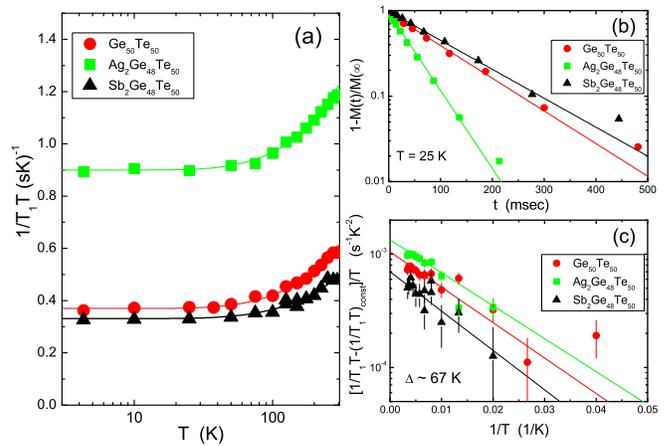}
\caption{(Color online)  (a) Temperature dependence of $^{125}$Te 1/$T_1T$  for Ge$_{50}$Te$_{50}$ (red circles), Ag$_{2}$Ge$_{48}$Te$_{50}$ (green squares) and Sb$_{2}$Ge$_{48}$Te$_{50}$ (black triangles).   The solid lines are best fit with the equation of $1/T_1T = (1/T_1T)_{\rm const} + A T e^{-E_{\rm g}/k_{\rm B}T} $ for each sample. 
(b) Typical nuclear recovery curves for the three samples at $T$ = 25 K. 
(c) Semi-log plot of  ($1/T_1T -  (1/T_1T)_{\rm const})/T$  versus 1/$T$.  The solid lines are fitting results with $E_{\rm g}/k_{\rm B}$ = $67(4)$ K.
}
\label{fig:T1}
\end{figure}

  Figure~\ref{fig:T1}(a)  shows temperature dependence of 1/$T_1T$ for the three samples.
   $T_1$ values reported here were measured by the single saturation pulse method at the peak position of the NMR spectra. 
    As shown in Fig. ~\ref{fig:T1}(b), the nuclear recovery data can be fitted by a single exponential function $1-M(t)/M(\infty) =  e^ {-t/T_{1}}$, where $M(t)$ and $M(\infty)$ are the nuclear magnetization at time $t$ after the saturation and the equilibrium nuclear magnetization at $t$ $\rightarrow$ $\infty$, respectively. 
    Similar to the case of $K$,   1/$T_1T$s for all samples exhibit qualitatively the same behavior: 
1/$T_1T$ decreases slightly with decreasing temperature, then levels off at low temperatures. 
     The temperature dependence of 1/$T_1T$ also can be explained by the two band model.

        In this case, 1/$T_1T$ is given by
\begin{eqnarray}
1/T_1T = (1/T_1T)_{\rm const} + A T e^{-E_{\rm g}/k_{\rm B}T} 
\label{eqn:T1}
\end{eqnarray}
where $(1/T_1T)_{\rm const}$ is the temperature independent constant value originated from the conduction carriers and the second term is due to thermal excitation effects from the second band. \cite{Blembergen1954,twobandmodel}
   A similar analysis of the temperature dependence of 1/$T_1T$ has been reported in the semimetal CaAl$_{2-x}$Si$_{2-x}$ (Ref. \onlinecite{Lue2007}) and the Heusler-type compound Fe$_{2+x}$V$_{1-x}$Al (Ref. \onlinecite{Kiyoshi2000}). 
   Using Eq.~\ref{eqn:T1}, the magnitude of $E_{\rm g}$ is estimated to be $67(4)$ K for Ge$_{50}$Te$_{50}$ and Ag$_{2}$Ge$_{48}$Te$_{50}$,  although the experimental data are somewhat scattered,
 as shown in Fig.~\ref{fig:T1}(c) where  $[(1/T_1T)-(1/T_1T)_{\rm const}]/T$ is plotted against to 1/$T$ on a semi-log scale.
    It is difficult to estimate $E_{\rm g}$ for Sb$_{2}$Ge$_{48}$Te$_{50}$ due to a large scattering of the data. 
   The black solid line in the figure is the best fit with a assumption of $E_{\rm g}/k_{\rm B}$ = $67$ K, which seems to reproduce the data reasonably although we cannot determine  $E_{\rm g}$. 
     It is noted that $67$ K is too small to attribute to the semiconducting gap energy of $0.3-0.5$ eV reported from optical measurements for GeTe.\cite{Nikolic1969}

\begin{figure}[b]
\centering
\centering
\includegraphics[width=\columnwidth]{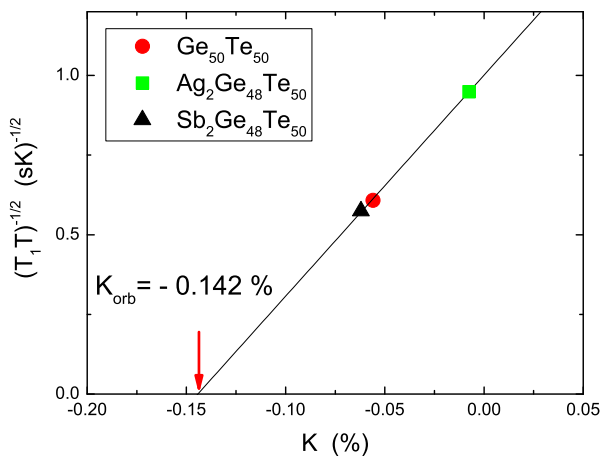}
\caption{(Color online) $(1/T_1T)_{\rm const}^{0.5}$ versus the temperature independent $K$ = $K_{\rm Pauli}$ + $K_{\rm orb}$ for three samples. The solid line is a linear fit giving rise to $K_{\rm orb}= -0.142\%$.}
\label{fig:Korb}
\end{figure}

      The solid lines in Fig.~\ref{fig:T1}(a)  are best fit to Eq.~\ref{eqn:T1}, using $E_{\rm g}/k_{\rm B}$ = $67$ K, with  $(1/T_1T)_{\rm const} = 0.37$ (sK)$^{-1}$,  $A$ = 0.0013 for Ge$_{50}$Te$_{50}$, $(1/T_1T)_{\rm const} = 0.90$ (sK)$^{-1}$,  $A$ = 0.0010 for Ag$_2$Ge$_{48}$Te$_{50}$, and $(1/T_1T)_{\rm const} = 0.33$ (sK)$^{-1}$,  $A$ = 0.00068 for Sb$_2$Ge$_{48}$Te$_{50}$, respectively.
      Within a Fermi liquid picture, $(1/T_1T)_{\rm const}$ is proportional to the square of the density of states at the Fermi level ${\cal N}(E_{\rm F}$) and $K_{\rm Pauli}$ is proportional to ${\cal N}(E_{\rm F}$). 
   Therefore, as $K_{\rm Pauli}$ is expected to be proportional to $(1/T_1T)_{\rm const}^{1/2}$, one can estimate the temperature independent $K_{\rm orb}$ by plotting $( 1/T_1T)_{\rm const}^{1/2}$ as a function of the temperature-independent $K$ = $K_{\rm Pauli}$ + $K_{\rm orb}$ at low temperatures for different samples.  
    As shown in Fig.~\ref{fig:Korb}, we actually found a linear relation between  $( 1/T_1T)_{\rm const}^{1/2}$ and  $K$ in the plot of $( 1/T_1T)_{\rm const}^{1/2}$ vs. the temperature independent $K$, from which $K_{\rm orb}$ is estimated to be $-0.142 \%$.

\begin{figure}[t]
\centering
\includegraphics[width=9.0 cm]{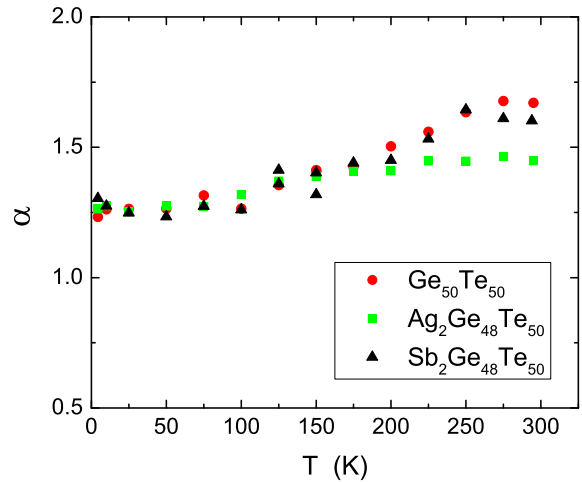}
\caption{(Color online)  Temperature dependence of the Korringa ratio $\alpha$ for Ge$_{50}$Te$_{50}$ (red circles), Ag$_{2}$Ge$_{48}$Te$_{50}$ (green squares) and Sb$_{2}$Ge$_{48}$Te$_{50}$ (black triangles)}
\label{fig:Korringa}
\end{figure}

     Using the NMR data, we can discuss electron correlations through the Korringa ratio analysis.
     As described,  both $(1/T_1T)_{\rm const}$  and $K_{\rm Pauli}$ are determined primarily by ${\cal N}(E_{\rm F}$). 
   This leads to the general Korringa relation  
\begin{eqnarray}
T_1TK_{\text{spin}}^2  = \frac{\hbar}{4\pi k_{\rm B}}\left(\frac{\gamma_{\rm e}}{\gamma_{\rm N}}\right)^2 \equiv R
\label{eqn:Korringa}
\end{eqnarray}
where $K_{\rm spin}$ denotes the spin part of the NMR shift.  
For the $^{125}$Te nucleus, ${R} =2.637\times 10^{-6}$ Ks. 
    Deviations from $R$ can reveal information about electron correlations in materials, which are conveniently expressed via the Korringa ratio $\alpha\equiv R/(T_1TK_{\text{spin}}^2)$.\cite{Moriya1963,Narath1968}
    For uncorrelated metals, one has $\alpha\sim1$ .  
    For antiferromagnetic spin correlation,  $\alpha >>1$; in contrast, $\alpha <<1$ for ferromagnetic spin correlations.
    The Korringa ratio $\alpha$, then, reveals how electrons correlate in materials. 
    Figure \ref{fig:Korringa} shows the temperature dependence of $\alpha$ for the three samples. 
    We found the values of $\alpha$ for all samples are similar, $\alpha$ $\sim$  1.25 at low temperatures, where the temperature independent (1/$T_1T)_{\rm const}$ and $K_{\rm Pauli}$  dominate,  indicative of no strong correlations for conduction carriers originated from self-doping/substitution effects in the samples. 
     With increasing temperature, $\alpha$ slightly increases above $\sim$ 50 K.
    If we assume that the Korringa relation holds at high temperatures, the increase suggests a tiny enhancement of antiferromagnetic spin correlations for carriers. 
     Since the temperature dependence of $\alpha$ originates from the second band having semiconducting nature, these results may suggest that thermally excited carriers play an important role in electron correlation effects in the system.
    As electron correlations have been pointed out to be significant for a figure of merit ($zT$ values),\cite{Joura2004}    
    it is interesting if the increase of $zT$ in Ge$_{50}$Te$_{50}$ at high temperatures above 300 K\cite{Levin2013} is related to the electron correlations. 
     Further studies at high temperature NMR measurements are required in shedding light on this issue. 

      Now we discuss how the carrier concentration changes by Ag or Sb substitution based on ${\cal N}(E_{\rm F}$) obtained from NMR data. 
     In a parabolic band for noninteracting carriers, ${\cal N}(E_{\rm F}$) is given by 
${\cal N}(E_{\rm F}$) = $\frac{4\pi}{h^3}(2m^*)^{3/2}E_{\rm F}^{1/2}$ where 
$E_{\rm F}^{1/2} = \frac{h^2}{2m^*}(3\pi^2n)^{2/3}$.
     Here $n$ is the carrier concentration and $m^*$ the renormalized effective carrier mass. 
     Therefore  one can get a simple relation of  ${\cal N}(E_{\rm F}$) $\propto$ ($m^*n^{1/3}$).
     From the values of $(1/T_1T)_{\rm const}$ and/or $K_{\rm Pauli}$ where the effect from $m^*$ can be negligible, the carrier concentration in Ag$_{2}$Ge$_{48}$Te$_{50}$ is found to increase about 380$\%$  from that of Ge$_{50}$Te$_{50}$ while the carrier concentration in Sb$_{2}$Ge$_{48}$Te$_{50}$ is reduced only by $\sim$ 16$\%$. 
     Since there are 1.85$\times10^{22}$ cm$^{-3}$ Ge atoms in Ge$_{50}$Te$_{50}$, the replacement of two Ag  atoms for  two Ge atoms out of 50 provides additional  7.4$\times10^{20}$ cm$^{-3}$ holes into the system. 
    On the other hand, the substitution of two Sb atoms should reduce the same amount of carrier concentration (7.4$\times10^{20}$ cm$^{-3}$).
    Therefore the large increase of the carrier concentration by the Ag substitution and the slight decrease of that by the Sb substitution  cannot be explained by the simple substitution effect. 
    Thus these results strongly indicate that the number of Ge vacancies must be different for Ag or Sb substitutions.
    A similar conclusion has been pointed out in our previous paper.\cite{Levin2016}

\begin{figure*}[tb]
\includegraphics[width=16.5 cm]{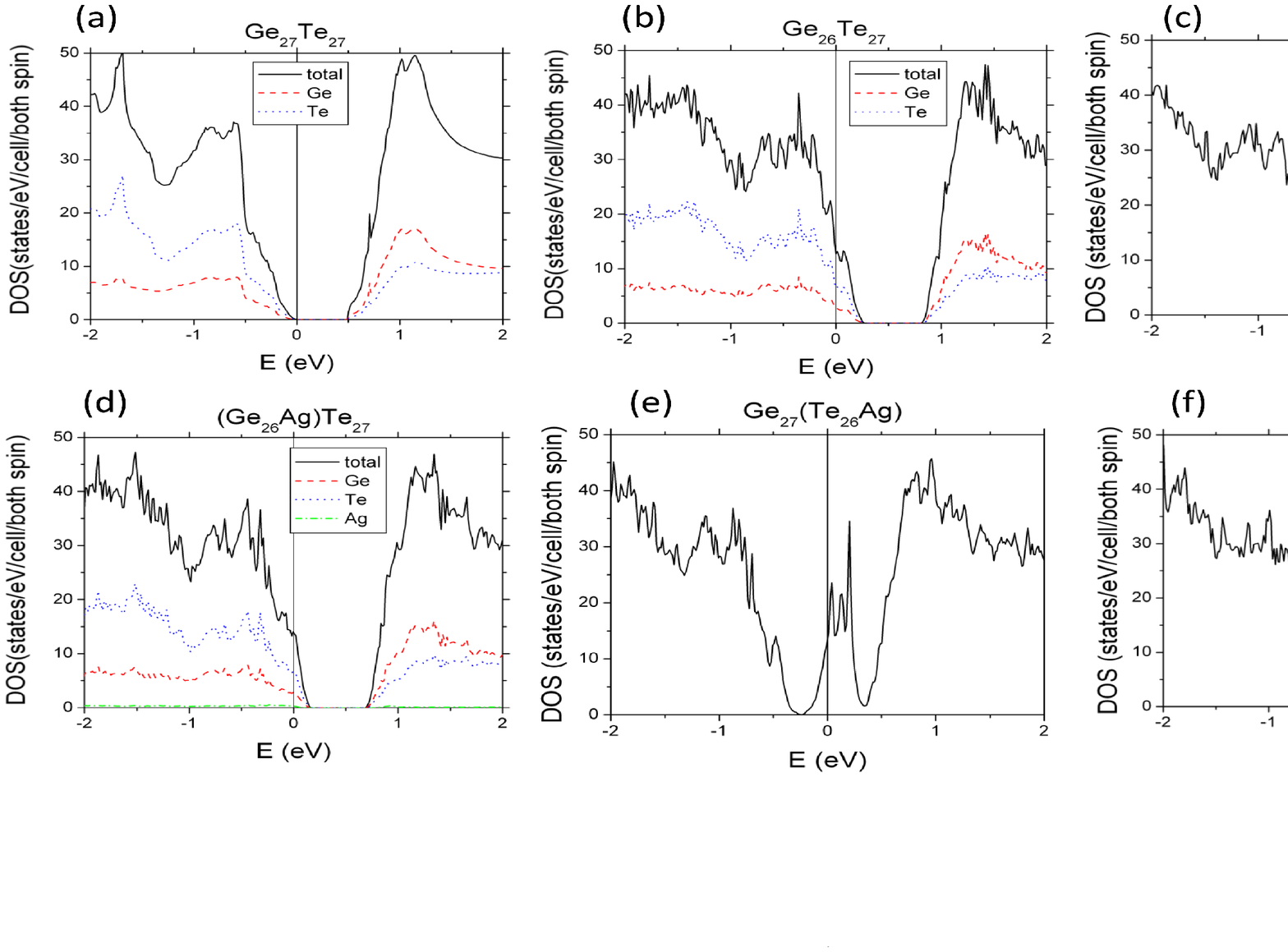}
\caption{(Color online) Density of states (DOS) near the Fermi level ($E_{\rm F}$). Black line is the total DOS. In some figures, atomic decomposed DOS is shown where the blue and red dotted lines show DOS from Te 5$p$ and Ge 4$p$ electrons, respectively: (a) Ge$_{27}$Te$_{27}$, (b) Ge$_{26}$Te$_{27}$, (c) Ge$_{27}$Te$_{26}$, (d) (Ge$_{26}$Ag)Te$_{27}$, (e) Ge$_{27}$(Te$_{26}$Ag), and (f) (Ge$_{26}$Ag)Te$_{26}$.    }
\label{fig:DOS1}
\end{figure*}


    To get insight of the origin of the metallic conductivity in Ge$_{50}$Te$_{50}$, particularly the vacancies effects on electronic structure of Ge$_{50}$Te$_{50}$, we performed first principles calculations where we employed a full-potential linear augmented plane wave method (FP-LAPW)\cite{Blaha2001} with a generalized gradient approximation (GGA) functional.\cite{Perdew1996}  
    We constructed supercells which are composed of 27 Ge atoms and 27 Te atoms and randomly choose sites for vacancies or for Ag substituted site.  
   For obtaining self-consistent charge density,  we employed $R_{\rm MT}k_{\rm max}$ = 7.0 and $R_{\rm MT}$ = 2.3 and 2.8 a.u. for Ge and Te atoms respectively.  
   We selected 828 ${\bf k}$-points in Irreducible Brillouin Zone for obtaining self-consistent charge and density of states (DOS). 
   As convergent criteria, we used energy difference 0.0001 Ry/cell, charge difference 0.0001 e, and force difference 1.0 mRy/a.u. between self-consistent steps. 
    To get optimized structure, we relaxed atoms around vacancy or the substituted  atom so that forces on each atom are less than 2.0 mRy/a.u.  

    Figure \ref{fig:DOS1}(a) shows the calculated DOS for a perfect Ge$_{27}$Te$_{27}$ without any defect, with a band gap of $\sim$ 0.5 eV (semiconductor nature). 
    This agrees well with previous reports.\cite{Polatoglou1982, Singh2013, Gevacancy2006}
     Here we show atomic decomposed DOS of the perfect Ge$_{27}$Te$_{27}$, where the black line shows the total DOS. 
     The red and blue dotted lines show DOS from Te 5$p$ and Ge 4$p$ electrons, respectively.
      Figures \ref{fig:DOS1}(b) and \ref{fig:DOS1}(c) show the vacancy effect on DOS. 
     In the case of a vacancy at the Ge site (Ge$_{26}$Te$_{27}$), Fermi level $E_{\rm F}$ moves to lower energy while keeping the similar gap structure with the case of  Ge$_{27}$Te$_{27}$. 
    This produces a finite DOS at $E_{\rm F}$, giving rise to metallic characters.
     Here the most part of DOS at $E_{\rm F}$ originate from Te 5$p$ and Ge 4$p$ electrons.
      On the other hand, a vacancy at the Te site (Ge$_{27}$Te$_{26}$) keeps semiconducting states although some isolated states are developed in the gap.
    Thus we conclude that a vacancy at the Ge site gives rise to $p$-type metallic conductivities in Ge$_{50}$Te$_{50}$ as has been observed in experiments. 
     A similar conclusion based on electronic structure calculations has been reported by Edwards ${\it et~al}$.\cite{Gevacancy2006}   
         We further investigate the Ag substitution effect on electronic states. 
    Figures \ref{fig:DOS1}(d) and \ref{fig:DOS1}(e) show a Ag atom substitution effect on DOS. 
     While replacing a Ge atom by a Ag atom lowers Fermi level and gives metallic characters as in the case of Ge vacancy, replacing a Te atom develops some isolated states in the gap and places $E_{\rm F}$  on the isolated states.
   Finally, Fig. \ref{fig:DOS1}(f)  shows DOS for a case that a Ag atom replaces a Ge atom and a vacancy on a the Te atom site. 
     In this case the impurity states are sharper than other cases and $E_{\rm F}$ is located at center of isolated states.
    As we discussed, our NMR data were well explained by the two band model where one band overlaps the Fermi level giving metallic nature and the other band is separated from the Fermi level by an energy gap of $E_{\rm g}/k_{\rm B}$ = $67(4)$ K.
    It is clear that  the metallic band can be attributed to the Ge vacancy effect, while the second band cannot be explained by the effect.   
    We found that a vacancy at the Te sites produces isolated state in the gap, and one  may think that it could be origin of the second band. 
    However, our observation of a gap magnitude of  $67(4)$ K [$5.8(3)$ meV] is much smaller than the gap energy of order (0.1 eV) even if we take the  isolated states created by the Te defects into consideration. 
   Therefore, we consider that the observed semiconducting nature cannot be attributed to the Te-defect effects but probably fine structures of DOS near the Fermi level.

\begin{figure}[tb]
\includegraphics[width=8.5cm]{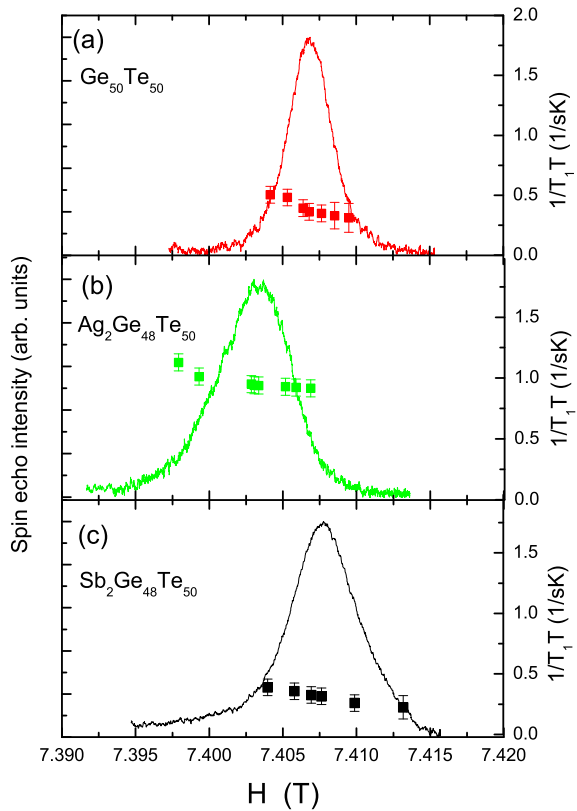}
\caption{(Color online) Position dependence of $1/T_1T$ at $T$ = 4.3 K for (a) Ge$_{50}$Te$_{50}$,  (b)  Ag$_{2}$Ge$_{48}$Te$_{50}$  and (c) Sb$_{2}$Ge$_{48}$Te$_{50}$,  together with corresponding NMR spectrum.
}
\label{fig:inhom}
\end{figure}

       Finally it is interesting to point out inhomogeneity of electronic states in the samples.
       According to Levin ${\it et~al.},$\cite{Levin2013PRB} electronic inhomogeneity have been observed in some semiconductors such as PbTe from $1/T_1$ measurements. 
      Here we investigate homogeneity of electronic states in the samples by measuring $T_1$ at $T$ = 4.3 K and different positions in the spectra for the three samples.
      As shown in Figs. 7 (a)-(c), $1/T_1$ seems to depend on the positions of the spectra, where we plotted 1/$T_1T$ together with the corresponding spectrum. 
     One can see that $1/T_1$ has a trend of the slight increase at lower magnetic field positions giving to greater $K_{\rm spin}$. 
      For example, the $1/T_1T$ at the peak position of Ge$_{50}$Te$_{50}$ is  $\sim$ 0.36 (sK)$^{-1}$, while the $1/T_1T$  $\sim$ 0.50 (sK)$^{-1}$at a lower field position ($H$ = 7.4033 T).
     The enhancement of 1/$T_1T$ and the larger $K_{\rm spin}$ at lower magnetic fields are consistent with an increased carrier concentration. 
    Since 1/$T_1T$ and $K_{\rm spin}$ values are related to ${\cal N}(E_{\rm F}$),  this result indicates that the electronic state in  Ge$_{50}$Te$_{50}$ is  likely inhomogeneous. 
    Similar behaviors have been observed in Ag$_2$Ge$_{48}$Te$_{50}$ and Sb$_2$Ge$_{48}$Te$_{50}$. 
     These results indicate electronic states of all GeTe-based materials investigated here are inhomogeneous which could originate from a possible inhomogeneous distribution of defects creating areas with differing carrier concentrations  in Ge$_{50}$Te$_{50}$ and M$_{2}$Ge$_{48}$Te$_{50}$ (M = Ag, Sb).

\section{Conclusion}
  We have carried out $^{125}$Te NMR measurements to microscopically investigate the electronic properties of Ge$_{50}$Te$_{50}$, Ag$_{2}$Ge$_{48}$Te$_{50}$ and Sb$_{2}$Ge$_{48}$Te$_{50}$.
   For Ge$_{50}$Te$_{50}$, NMR shift $K$ and 1/$T_1T$ data are nearly temperature independent at low temperatures below $\sim$ 50 K and both increase slightly with increasing temperature at high temperatures. 
   These behavior are well explained by a two band model where one band overlaps the Fermi level and the other band is separated from the Fermi level by an energy gap of $E_{\rm g}/k_{\rm B}$ = $67(4)$ K.
   Korringa analysis indicates that the conduction carriers can be considered as free carriers with  no significant electron correlations at low temperatures.
   On the other hand, Korringa ratio increases slightly at high temperature, suggesting  the slight enhancement of the electron correlation.  
   First principle calculation revealed that the metallic band originates from the Ge vacancy while the semiconductor-like band may be related to the fine structure of the density of states near the Fermi level. 
    Low temperature $^{125}$Te NMR data for Ag$_{2}$Ge$_{48}$Te$_{50}$ and Sb$_{2}$Ge$_{48}$Te$_{50}$ clearly demonstrate that the carrier concentration changes by  Ag or Sb substitutions where the Ag substitution increases  hole concentration while Sb substitution decreases the concentration.

\begin{acknowledgments}
The research was supported by the U.S. Department of Energy, Office of Basic Energy Sciences, Division of Materials Sciences and Engineering. Ames Laboratory is operated for the U.S. Department of Energy by Iowa State University under Contract No.~DE-AC02-07CH11358. 
\end{acknowledgments}

\end{document}